\documentstyle[amssymb,twocolumn,prb,aps,epsf]{revtex}

\def\be{\begin{equation}}
\def\ee{\end{equation}}
\def\bea{\begin{eqnarray}}
\def\eea{\end{eqnarray}}

\begin{document}

\twocolumn[\hsize\textwidth\columnwidth\hsize\csname
@twocolumnfalse\endcsname

\title{Kondo and anti-Kondo resonances in transport through nanoscale devices}
\author{A. A. Aligia and C. R. Proetto}
\address{Comisi\'{o}n Nacional de Energ\'{\i}a At\'{o}mica\\
Centro At\'{o}mico Bariloche and Instituto Balseiro\\
8400 Bariloche, Argentina}
\maketitle

\begin{abstract}
We study the current through a quantum wire side coupled to a quantum dot,
and compare it with the case of an embedded dot. The system is modeled by
the Anderson Hamiltonian for a linear chain, with one atom either coupled to
(side-dot) or substituted by (embedded dot) a magnetic impurity. For
realistic (small) hopping of the dot to the rest of the system, an exact
relationship between both conductivities holds. We calculate the temperature
dependence for moderate values of the Coulomb repulsion $U$ using an
interpolative perturbative scheme. For sufficiently large $U$ and
temperature greater than the Kondo temperature, the conductance as a
function of gate voltage displays two extrema.
\end{abstract}


]
\narrowtext

\section{Introduction }

Transport through quantum dots have been studied in a series of recent
experiments,\cite{gold1,cro,gold2} in which the peculiar features of the
spectral density in the Kondo effect are manifested. An important fact in
these experiments is that only one Kondo impurity is present.

While these experiments correspond to embedded quantum dots, very recent
studies provide predictions for other situations, in particular a quantum
dot side coupled to a quantum wire.\cite{bul,kang,torio} In Refs. 4 and 5,
decoupling approximations were used for infinite Coulomb repulsion $U,$
which have limitations at finite temperatures.\cite{hew} For example, in the
slave-boson mean-field approximation,\cite{kang} there is an unphysical
phase transition at which the impurity decouples from the rest of the
system, either when the temperature $T=T_{K}$ (Kondo temperature) or for $%
\varepsilon _{F}-\varepsilon _{d}\gg \Delta $, where $\varepsilon _{F}$ is
the Fermi energy, $\varepsilon _{d}$ is the impurity level and $\Delta
=\Gamma /2$ is the half-width half-maximum of the resonant level.\cite{fran}
The method used by Torio {\it et al.} is expected to be very accurate,\cite
{torio,ferr} but it is limited to $T=0.$

Here we report calculations of the conductance as a function of temperature
and gate voltage for embedded and quantum dots for $U\leq 8\Delta .$ In some
experimental situations $\Delta \sim 0.15$ meV, $U\sim 0.6$ meV.\cite
{gold1,gold2} Thus, this work complements previous ones for infinite $U$ or $%
T=0$, and presents a more systematic study of the temperature dependence for
moderate $U$. We use second-order perturbation theory in $U$\cite
{yos,hor,sal} modified to ensure that the approximation is exact in
different limits, including $U\rightarrow +\infty .$\cite{fer,levy,kaju}

The paper is organized as follows. In section II we write the relevant
expressions which relate the conductance with the spectral density at the
dot. The approximation for the latter is briefly reviewed and discussed in
section III. Section IV contains the main results. Section V contains a
brief summary and discussion.

\section{Conductance for embedded and side dots}

For both cases, the Hamiltonian can be written as an Anderson model 
\begin{equation}
H=H_{0}+H_{con},  \label{h}
\end{equation}
where $H_{0}$ describes a tight-binding chain without site $0$ plus a
disconnected quantum dot, and $H_{con}$ contains the hopping terms which
connect the quantum dot with the chain, 
\bea
H_{0}=-t\sum_{\sigma ,i\neq 0,1}\left( c_{i\sigma }^{\dagger }c_{i-1\sigma
}+H.c.\right) +\varepsilon _{d}\sum_{\sigma }d_{\sigma }^{\dagger }d_{\sigma
} \nonumber 
\\
+ Ud_{\uparrow }^{\dagger }d_{\uparrow }d_{\downarrow }^{\dagger
}d_{\downarrow }.  \label{h0}
\eea
For the embedded dot, 
\begin{equation}
H_{con}^{e}=-\sum_{\sigma }\left[ d_{\sigma }^{\dagger }\left(
t_{-1}^{\prime }c_{-1\sigma }+t_{1}^{\prime }c_{1\sigma }\right)
+H.c.\right] ,  \label{he}
\end{equation}
while for the side dot, 
\begin{equation}
H_{con}^{s}=-\sum_{\sigma }\left[ c_{0\sigma }^{\dagger }\left(
t_{-1}c_{-1\sigma }+t_{1}c_{1\sigma }+t^{\prime }d_{\sigma }\right)
+H.c.\right] .  \label{hs}
\end{equation}

The conductance for an embedded dot can be written as\cite{meir} 
\begin{equation}
G_{e}=\frac{2\pi e^{2}}{h}\sum_{\sigma }\int \left( -\frac{df}{d\omega }%
\right) \Gamma _{\sigma }(\omega )\rho _{d\sigma }(\omega ),  \label{ge}
\end{equation}
where $f(\omega )$ is the Fermi function, $\rho _{d\sigma }(\omega )$ is the
spectral density of the $d_{\sigma }$ states, and 
\begin{equation}
\Gamma _{\sigma }(\omega )=\frac{\Gamma _{-1\sigma }(\omega )\Gamma
_{1\sigma }(\omega )}{\Gamma _{-1\sigma }(\omega )+\Gamma _{1\sigma }(\omega
)},\text{ with }\Gamma _{i\sigma }(\omega )=2\pi \left( t_{i}^{\prime
}\right) ^{2}\rho _{i\sigma }(\omega )  \label{gam}
\end{equation}
where $\rho _{i\sigma }(\omega )$ is the spectral density of the states
described by the operators $c_{i\sigma }$ in $H_{0}.$ From the equations of
motion for a semi-infinite chain, one easily obtains $\rho _{-1\sigma }=\rho
_{1\sigma }=\left[ 1-(\omega /2t)^{2}\right] ^{1/2}/\pi t.$ The resonant
level width of the $d$ states is $\Delta _{\sigma }(\omega )=\pi
\sum\limits_{i}\left( t_{i}^{\prime }\right) ^{2}\rho _{i\sigma }(\omega ).$
In the following we assume that $t_{i}^{\prime }$ is small, implying that
the Kondo temperature is much smaller than the band width. The sum over spin
indices reduces to a factor 2, since for zero magnetic field there is no
spin dependence. Then, 
\begin{equation}
G_{e}=\frac{2e^{2}}{h}\frac{4\pi \Delta _{e}\left( t_{1}^{\prime
}t_{-1}^{\prime }\right) ^{2}}{\left[ (t_{1}^{\prime })^{2}+(t_{-1}^{\prime
})^{2}\right] }\int \left( -\frac{df}{d\omega }\right) \rho _{d\sigma
}(\omega ),  \label{ge2}
\end{equation}
with 
\begin{equation}
\Delta _{e}=\frac{(t_{1}^{\prime })^{2}+(t_{-1}^{\prime })^{2}}{t}\sqrt{%
1-\left( \frac{\varepsilon _{F}}{2t}\right) ^{2},}  \label{dele}
\end{equation}
where $\varepsilon _{F}$ is the Fermi energy. $\Delta _{e}$ is the
half-width at half-maximum of $\rho _{d\sigma }(\omega )$ for $U=0.$ Both $%
\Delta _{e}$ and $U$ are the parameters of the perturbation calculation,
which allows to obtain $\rho _{d\sigma }(\omega )$ as described in the next
section.

For the side dot, Eqs. (\ref{ge},\ref{ge2}) apply with the substitution $%
\rho _{d\sigma }\rightarrow \rho _{0\sigma },$ and $t_{i}^{\prime
}\rightarrow t.$ Assuming $t_{-1}=t_{1}=t$ (periodic chain) for simplicity
and $t^{\prime }\ll t$ as before, we have 
\begin{equation}
G_{s}=\frac{2e^{2}}{h\rho (\varepsilon _{F})}\int \left( -\frac{df}{d\omega }%
\right) \rho _{0\sigma }(\omega ),  \label{gs}
\end{equation}
where $\rho (\omega )=1/\left[ \pi \left( 4t^{2}-\omega ^{2}\right)
^{1/2}\right] $ is the density of states of any site of the chain for given
spin and $t^{\prime }=0.$ The corresponding density of states at site $0$
for $t^{\prime }\neq 0$ ($\rho _{0\sigma }(\omega )$) can be expressed in
terms of $\rho _{d\sigma }(\omega ).$ Using equations of motion for the
Green's functions, one can write: 
\begin{equation}
G_{0\sigma }(t^{\prime },\omega )=G_{0\sigma }(0,\omega )+(t^{\prime
})^{2}G_{d\sigma }(\omega )G_{0\sigma }^{2}(\omega ),  \label{g0}
\end{equation}
where $G_{0\sigma }(t^{\prime },\omega )=\langle \langle c_{0\sigma
};c_{0\sigma }^{\dagger }\rangle \rangle _{\omega }$ is the Green function
for site $0$ and hopping $t$ with the dot. Similarly $G_{d\sigma }(t^{\prime
},\omega )=\langle \langle d_{\sigma };d_{\sigma }^{\dagger }\rangle \rangle
_{\omega }$ is the Green function for the electrons in the dot. From Eq. (%
\ref{g0}) and $G_{0\sigma }(0,\omega )=-i\pi \rho (\omega ),$one has 
\begin{equation}
\rho _{0\sigma }(\omega )=-\frac{1}{\pi }%
\mathop{\rm Im}%
G_{0\sigma }(t^{\prime },\omega )=\rho (\omega )-\left[ \pi t^{\prime }\rho
(\omega )\right] ^{2}\rho _{d\sigma }(\omega ).  \label{ro0}
\end{equation}
Replacing Eq. (\ref{gs}) and neglecting again the dependence of $\rho
(\omega )$ on $\omega $ near $\varepsilon _{F},$we can write 
\begin{equation}
G_{s}=\frac{2e^{2}}{h}\left[ 1-\pi \Delta _{s}\int \left( -\frac{df}{d\omega 
}\right) \rho _{d\sigma }(\omega )\right] ,  \label{gs2}
\end{equation}
where 
\begin{equation}
\Delta _{s}=\frac{(t^{\prime })^{2}}{\sqrt{4t^{2}-\varepsilon _{F}^{2}}}.
\label{dels}
\end{equation}
We see that in both cases, $G_{e}$ and $G_{s}$ are determined by an integral
of the same form, involving the density of states at the dot $\rho _{d\sigma
}(\omega )$, but with different sign. An increase in $\rho _{d\sigma
}(\omega )$ for an embedded (side) dot leads to an increase (decrease) in $%
G_{e}$ $\left( G_{s}\right) .$

At $T=0$, the conductivities can be related with the occupation of the
impurity level $n_{d\sigma }=\langle d_{\sigma }^{\dagger }d_{\sigma
}\rangle $ , using Friedel's sum rule.\cite{lan} For $\varepsilon _{eff}$
and $\Delta $ independent of frequency, this rule states that 
\begin{equation}
\rho _{d\sigma }(\varepsilon _{F})=\frac{\sin ^{2}(\pi n_{d\sigma })}{\pi
\Delta },  \label{sum}
\end{equation}
and then, replacing in (\ref{ge2}) and (\ref{gs2}) one obtains the simple
result,

\[
G_{e}=\frac{2e^{2}}{h}\frac{4\left( t_{-1}^{\prime }t_{1}^{\prime }\right)
^{2}}{\left[ (t_{1}^{\prime })^{2}+(t_{-1}^{\prime })^{2}\right] ^{2}}\sin
^{2}(\pi n_{d\sigma }),
\]
\begin{equation}
G_{s}=\frac{2e^{2}}{h}\cos ^{2}(\pi n_{d\sigma }).  \label{gsum}
\end{equation}
Thus, the ideal conductance for an embedded dot is obtained for a symmetric
dot $t_{-1}^{\prime }=t_{1}^{\prime }$ and when $V_{g}$ is tuned to the
symmetric Anderson model $(\varepsilon _{eff}=\varepsilon _{F}).$ For the
side dot, the dependence with $V_{g}$ is the opposite, as noted earlier.\cite
{kang,torio}

\section{The spectral density}

The dot Green's function can be written in the form: 
\begin{equation}
G_{d\sigma }(\omega )=\frac{1}{\omega -\varepsilon _{eff}+i\Delta -\Sigma
_{\sigma }(\omega ,T)}.  \label{gp}
\end{equation}
For the embedded dot, $\Delta $ is given by Eq. (\ref{dele}) and 
\begin{equation}
\varepsilon _{eff}^{e}=\varepsilon _{d}-eV_{g}+Un_{d\overline{\sigma }}-%
\frac{\varepsilon _{F}}{2t^{2}}\left[ (t_{1}^{\prime })^{2}+(t_{-1}^{\prime
})^{2}\right] ,  \label{effe}
\end{equation}
where the subscript $\sigma $ in $\varepsilon _{eff}^{e}$ has been dropped
for simplicity and $V_{g}$ is a gate voltage which controls the dot level.
The dependence on frequency of the terms proportional to $\left(
t_{i}^{\prime }\right) ^{2}$ has been neglected assuming $t^{\prime }\ll t$
as before. For the side dot, $\Delta $ was written in Eq. (\ref{dels}) and 
\begin{equation}
\varepsilon _{eff}^{s}=\varepsilon _{d}-eV_{g}+Un_{d\overline{\sigma }.}
\label{effs}
\end{equation}

In traditional second-order perturbation theory in $U,$ the self-energy $%
\Sigma _{\sigma }(\omega ,T)$ is calculated from a Feynmann diagram which
involves two sums in Matsubara frequencies:\cite{yos,hor,sal} 
\bea
\Sigma _{\sigma }^{(2)}(i\omega _{l},T) &=&-(UT)^{2}\sum_{n,m}G_{d\sigma
}^{0}(i\omega _{l}-i\nu _{m})G_{d\bar{\sigma}}^{0}(i\omega _{n})  \nonumber
\\
&&\times G_{d\bar{\sigma}}^{0}(i\omega _{n}+i\nu _{m}),  \label{self}
\eea
where the unperturbed Green's function is:

\begin{equation}
G_{d\sigma }(\omega )=\frac{1}{\omega -\varepsilon _{eff}^{0}+i\Delta }.
\label{g00}
\end{equation}
The choice of $\varepsilon _{eff}^{0}$ is arbitrary and is equivalent to the
division of the Hamiltonian between unperturbed part and perturbation.
Taking $\varepsilon _{eff}^{0}=\varepsilon _{eff}$ is equivalent to a sum of
an infinite series of diagrams.\cite{hor,sal} If the expectation value $%
n_{d\sigma }$ were calculated using $G_{d\sigma }^{0},$ $\varepsilon _{eff}$
would represent the Hartree-Fock effective $d$ level. However, we find that
in general calculating $n_{d\sigma }$ with $G_{d\sigma }(\omega )$ agrees
better with Friedel's sum rule.\cite{lan} Then 
\begin{equation}
n_{d\sigma }=\int f(\omega )\rho _{d\sigma }(\omega )d\omega ,\text{ with }%
\rho _{d\sigma }(\omega )=-\frac{1}{\pi }%
\mathop{\rm Im}%
G_{d\sigma }(\omega ).  \label{nd}
\end{equation}
$n_{d\sigma }$ should be determined self-consistently from Eqs. (\ref{nd},%
\ref{self}), and (\ref{effe}) or (\ref{effs}), for each value of $V_{g}.$

\begin{figure}
\narrowtext
\epsfxsize=3.5truein
\vbox{\hskip 0.05truein \epsffile{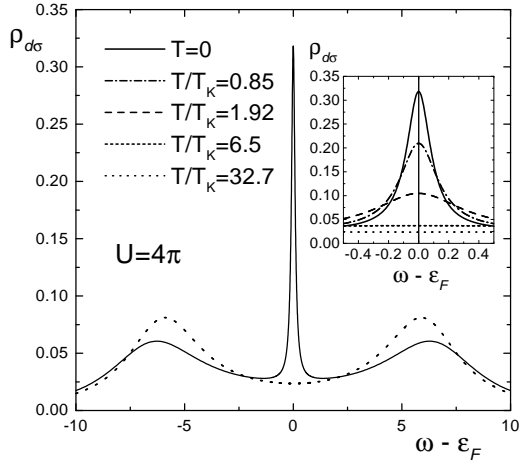}}
\medskip
\caption{Spectral density at the dot for given spin, $U=4\pi ,$ $\varepsilon
_{d}=-2\pi ,$ $\varepsilon _{F}=0 ,$ and several temperatures. $\Delta =1$
is chosen as the unit of energy.}
\end{figure}

A particular situation is the case $\varepsilon _{eff}=\varepsilon _{F},$ at
which $\rho _{d\sigma }(\omega )$ is symmetric around $\varepsilon _{F}$ and
then $n_{d\sigma }=1/2.$ At this point, the perturbative result satisfies
exactly Friedel's sum rule.\cite{lan} Out of this point, for large $U$, the
perturbative result should be improved as discussed later in this section.
For the symmetric model $(\varepsilon _{eff}=\varepsilon _{F}),$ the
spectral density coincides with that obtained using Quantum Monte Carlo with
Maximum Entropy method within statistical errors if $U/\Delta \lesssim
1.25\pi \simeq 3.93.$\cite{sil} For $U/\Delta \simeq 7.6,$ the difference
between both results is $\sim 10\%.$ In Fig. 1 we show the spectral density
in the symmetric case for $U/\Delta =4\pi .$ Although this ratio is beyond
the validity of the perturbation theory, the result is in qualitative
agreement with the corresponding result obtained using Wilson's
Renormalization Group (WRG),\cite{cos} The central (or Kondo) peak is
broader in perturbation theory. However, if the Kondo temperature is defined
as half width at half maximum $(T_{K}\simeq 0.10\Delta ),$ the temperature
dependence of the Kondo peak for $T<2T_{K}$ agrees within $20\%$ with WRG
results. The temperature dependence of the rest of $\rho _{d\sigma }$ is
very small.

As $\varepsilon _{eff}$ increases from the symmetric situation $\varepsilon
_{eff}=\varepsilon _{F},$ the Kondo peak is slightly shifted to higher
energies, but $\rho _{d\sigma }(\varepsilon _{F})$ remains high. However,
for $U>5\Delta $, the perturbative result shifts to the opposite direction,
leading to a violation of Friedel's sum rule. This shortcoming can be cured
using an interpolative solution for the self-energy:\cite{fer,levy,kaju}

\begin{equation}
\Sigma _{\sigma }(\omega ,T)=\frac{n_{d\overline{\sigma }}(1-n_{d\overline{%
\sigma }})\Sigma _{\sigma }^{(2)}}{n_{d\overline{\sigma }}^{0}(1-n_{d%
\overline{\sigma }}^{0})-[(1-n_{d\overline{\sigma }})U+\varepsilon
_{d}^{V}-\varepsilon _{eff}^{0}]U^{-2}\Sigma _{\sigma }^{(2)}},  \label{smk}
\end{equation}
where $n_{d\sigma }^{0}=-\int f(\omega )%
\mathop{\rm Im}%
G_{d\sigma }^{0}(\omega )d\omega /\pi $ is the impurity occupation
corresponding to the unperturbed impurity spectral density, and $\varepsilon
_{d}^{V}=\varepsilon _{eff}-Un_{d\overline{\sigma }}$ is the effective
impurity level for $U=0$ (see Eqs. (\ref{effe}) and (\ref{effs})), which
depends on the applied gate voltage $V_{g}$. This self energy leads to the
exact $G_{d\sigma }(\omega )$ not only for $U=0$, but also for a decoupled
dot ($\Delta =0$), and reproduces the leading term for $\omega \rightarrow
\infty $. The effective unperturbed impurity level $\varepsilon _{eff}^{0}$
can be calculated selfconsistently at arbitrary temperatures imposing the
condition $n_{d\sigma }^{0}=n_{d\sigma }$ \cite{fer,levy}. Unless otherwise
stated, the results presented here are obtained following this approach. For 
$T=0$, the results can be improved further determining $\varepsilon
_{eff}^{0}$ by imposing Friedel's sum rule, as done by Kajueter and Kotliar
to solve the impurity problem in the dynamical mean-field approach.\cite
{kaju} We call $\Sigma _{KK}$ the self energy obtained in this way.

\begin{figure}
\narrowtext
\epsfxsize=3.5truein
\vbox{\hskip 0.05truein \epsffile{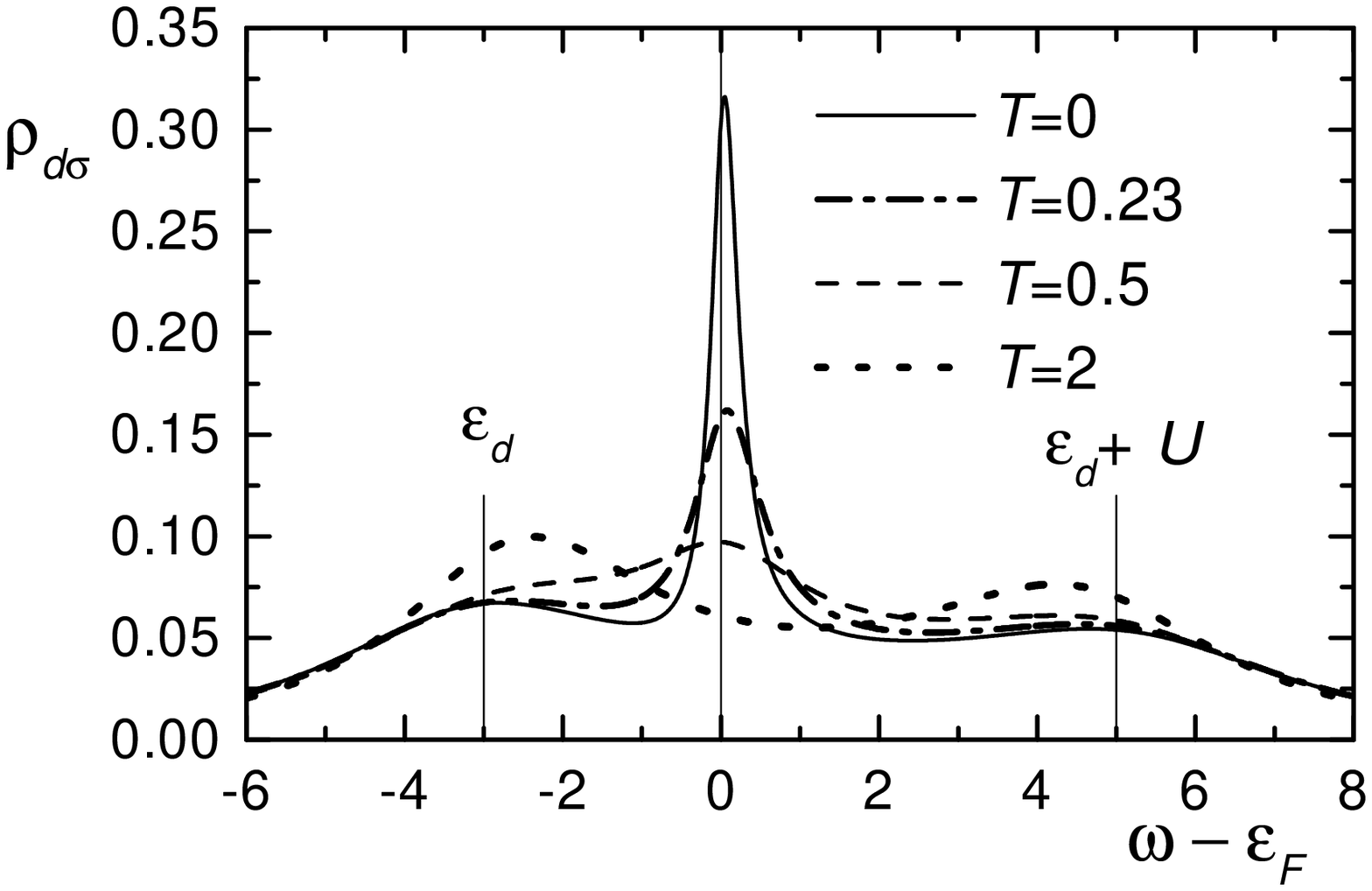}}
\medskip
\caption{}
\end{figure}

In Fig. 2 we show the spectral density for $U=8\Delta $ in an asymmetric
case with $\varepsilon _{d}-\varepsilon _{F}=-3\Delta $. The spectral
density shows the characteristic charge fluctuation peaks near $\varepsilon
_{d}$ and $\varepsilon _{d}+U,$ and the Kondo peak near $\varepsilon _{F}$.
The slave-boson mean-field theory \cite{kang} cannot reproduce the charge
fluctuation peaks and brings an incomplete description of the spectral
density. For the symmetric case $\varepsilon _{d}-\varepsilon _{F}=-4\Delta $%
, the Kondo temperature defined as half-width at half-maximum of the Kondo
peak is $T_{K}\simeq 0.23\Delta $.\cite{note} In a scale of $T_{K}$, for
increasing temperatures, the Kondo peak is rapidly suppressed, while the
charge fluctuation peaks remain and absorb the spectral weight of the Kondo
peak. For $T\sim T_{K}$, the density of states at the Fermi level $\rho
_{d\sigma }(\varepsilon _{F})$ is reduced to approximately half its value at 
$T=0$.

\begin{figure}
\narrowtext
\epsfxsize=3.5truein
\vbox{\hskip 0.05truein \epsffile{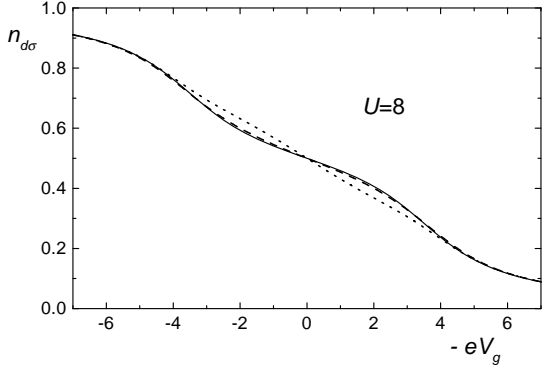}}
\medskip
\caption{Dashed line: dot occupation for a given spin as a function of gate
voltage for $U=8$, and $T=0$. Dotted line: corresponding result obtained
from the phase shift at the Fermi level using Friedel's sum rule.
\protect\cite{lan} Full line: result using $\Sigma _{KK}$ (imposing
Friedel's sum rule). The zero of $V_{g}$ is set at the point where $%
\varepsilon _{eff}=\varepsilon _{F},$ $n_{d\sigma }=1/2.$}
\end{figure}

In Fig. 3 we compare the resulting $n_{d\sigma }$ at $T=0$, with that
obtained from the phase shift \cite{lan} for the largest $U$ (worst case)
considered in this work ($U/\Delta =8$)$.$ We see that the deviations are
small and concentrated near but out of the symmetric case. Note that the
result using $\Sigma _{KK}$ (in which Friedel's sum rule is imposed) is very
similar to that obtained imposing $n_{d\sigma }^{0}=n_{d\sigma }$. In turn,
this result agrees very well with exact calculations using the Bethe ansatz.%
\cite{fer}

\section{Results for the conductance}

From Eqs. (\ref{ge2}), (\ref{dele}), (\ref{gs2}), (\ref{dels}), (\ref{gp}), (%
\ref{effe}), (\ref{effs}), and (\ref{nd}), one can see that for given $U$
and $t^{\prime }\left( \text{or }t_{i}^{\prime }\right) \ll t,$ the
conductance for the embedded and side dots depend on two parameters: $\Delta 
$ and $\varepsilon _{d}-eV_{g}.$ Furthermore, independently of the
approximation used for the spectral density, both conductances are related
shifting $V_{g}$ (or $\varepsilon _{d}$) and rescaling $\Delta :$%
\begin{equation}
G_{e}(\Delta ,V_{g}^{\prime })=\frac{4(t_{1}^{\prime }t_{-1}^{\prime })^{2}}{%
[(t_{1}^{\prime })^{2}+(t_{-1}^{\prime })^{2}]^{2}}\left[ \frac{2e^{2}}{h}%
-G_{s}(\Delta ,V_{g})\right] ,  \label{rel}
\end{equation}
with 
\[
V_{g}^{\prime }=V_{g}+\left[ (t_{1}^{\prime })^{2}+(t_{-1}^{\prime
})^{2}\right] \frac{\varepsilon _{F}}{2t^{2}e}.
\]
Thus, it is sufficient to calculate one of the conductances. 

\begin{figure}
\narrowtext
\epsfxsize=5.truein
\vbox{\hskip 0.05truein \epsffile{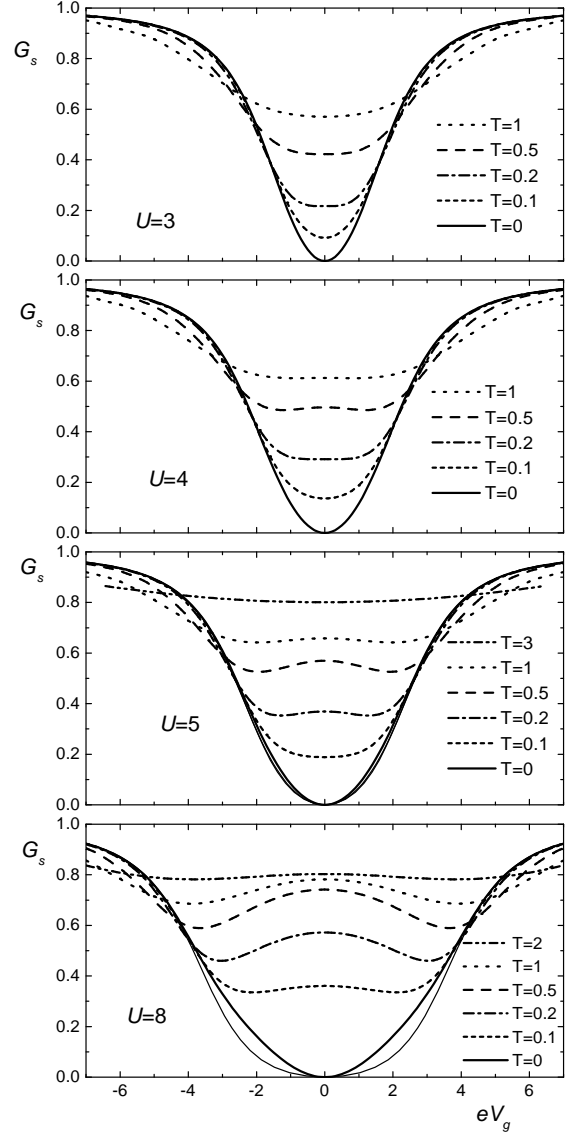}}
\medskip
\caption{Conductance for the side dot in units of $2e^{2}/h$ 
as a function of gate voltage for several
values of $U$ and $T$ indicated inside each figure. The bottom curve with
thiner full line was obtained using $\Sigma _{KK}$.}
\end{figure}

In Fig. 4 we
represent $G_{s}$ for several values of $U$ and $T.$ The unit of energy is
set as $\Delta =1,$ the conductance is expressed in units of $2e^{2}/h$ and
the zero of the gate voltage is set at the symmetric Anderson model ($%
eV_{g0}=\varepsilon _{d}+U/2=0$ for the side dot)$.$ For $T=0$ we have also
included the result imposing Friedel's sum rule,\cite{kaju} (using $\Sigma
_{KK}$) which gives a slightly smaller value of $G_{s}$ and allows to
estimate the error at finite $T$. This error is due to the effect on $\rho
_{d\sigma }(\varepsilon _{F})$ in the real part of $\Sigma (\varepsilon _{F})
$ for the asymmetric Anderson model. The imaginary part is always zero as it
should be for a Fermi liquid. For $U=8$, the difference in $G_{s}$ at $T=0$
between both interpolative methods attains its maximum value $\sim
0.07\times 2e^{2}/h$ near $eV_{g}-eV_{g0}=\pm 2.5$. For $U=5$, the maximum
deviations are of the order of 2\%, and for smaller $U$ they are negligible.
Note that the difference between both approaches is much smaller in integral
properties, like $n_{d\sigma }$ (see Fig. 3). As a consequence $G_{s}$
obtained using Eq. (\ref{gsum}) (instead of Eq. (\ref{gs2})) is practically
the same in both approaches. Using $\Sigma _{KK}$, of course Eq. (\ref{gs2})
and Eq. (\ref{gsum}) give identical results.

The conductance for $T=0$ and $U\geq 5$ are qualitatively similar to the
corresponding result of Ref. 6. As the temperature is increased, $G_{s}$
near $V_{g}=0$ rapidly increases. This is expected from the temperature
dependence of the Kondo peak, presented in the previous section. For $U\geq 5
$, at temperatures of the order of the Kondo temperature $(T_{K}\simeq 0.45$
for $U=5$ and $T_{K}\simeq 0.23$ for $U=8$ defined as half width at half
maximum of the Kondo spectral peak$),$ or slightly smaller, the conductance
shows two minima. For $T>T_{K}$, the resulting structure is qualitatively
what one expects for $t^{\prime }\rightarrow 0,$ in which only the Coulomb
blockade peaks at $\varepsilon _{d}$ and $\varepsilon _{d}+U$ are important
at finite temperatures. However, the peaks are displaced towards zero,
particularly for small $U$ and $T$. For $U=5$ and $T>T_{K}$, the
interpolative scheme and the ordinary second order perturbative result
(using $\Sigma _{\sigma }=\Sigma _{\sigma }^{(2)}$ and with selfconsistency
in $n_{d\sigma }$ only) are practically identical. At smaller temperatures,
the ordinary treatment exaggerates the structure with two minima. Instead,
the comparison with the results using $\Sigma _{KK}$ at $T=0$ and $U=8$,
suggest that the interpolative treatment with $n_{d\sigma }=n_{d\sigma }^{0}$
inhibits somewhat this structure for large $U.$

\begin{figure}
\narrowtext
\epsfxsize=3.5truein
\vbox{\hskip 0.05truein \epsffile{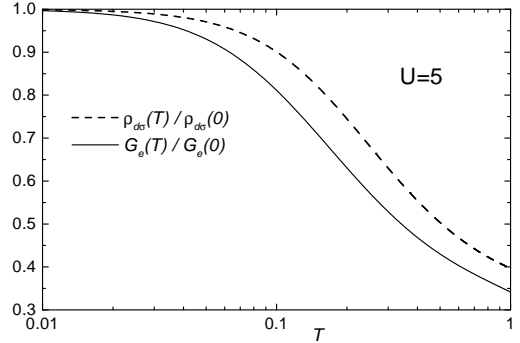}}
\medskip
\caption{Temperature dependence of the conductance for the embedded dot and
the density of states at the Fermi level for the symmetric Anderson model
and $U=5\Delta .$}
\end{figure}

As $U$ is decreased, the structure with two minima weakens. For $U=3\Delta $
the Anderson model is near the intermediate valence regime, since the Kondo
regime requires $U\gg \Delta .$ While the spectral density displays only one
peak, without marked shoulders for $U=3,$ plateaus reminiscent to the
transition to the Coulomb blockade regime are still present in the
conductance for $0.2\leq T\leq 0.5$. In Fig. 5 we show the temperature
dependence of the density of states at $\varepsilon _{F}$ and the
conductance of the embedded dot for $V_{g}=0$ and $U=5.$ The shape of both
quantities in a logarithmic temperature scale is similar to the
experimentally observed conductance.\cite{gold2} In our case the linear
behavior with $\log T$ extends over less than an order of magnitude: from $%
T\sim 0.13$ to $\sim 0.5$ for $\rho _{d\sigma }(\varepsilon _{F})$ and from $%
\sim 0.1$ to $0.3$ for $G_{e}.$ This corresponds to values lower or of the
order of $T_{K}.$

\begin{figure}
\narrowtext
\epsfxsize=3.5truein
\vbox{\hskip 0.05truein \epsffile{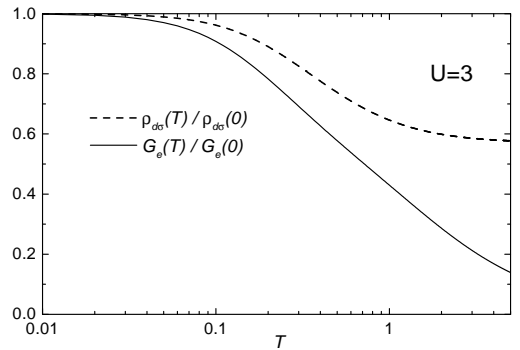}}
\medskip
\caption{Same as Fig. 6 for $U=3\Delta .$}
\end{figure}

In Fig. 6, we show the same temperature dependence for $U=3.$ In this case,
the half width at half maximum of the spectral density in the symmetric case
is $0.73.$ In comparison with Fig. 5, the behavior of $\rho _{d\sigma
}(\varepsilon _{F})$ and $G_{e}$ vs. $T$ is qualitatively different. $\rho
_{d\sigma }(\varepsilon _{F})\sim \log T$ for $0.2\lesssim T\lesssim 0.6.$
In contrast, except for a very small curvature upwards, $G_{e}$ is linear in 
$\log T$ for $0.16\lesssim T\lesssim 2.$ A quadratic fit of all points with
temperatures an integer times 0.01 in this interval gives $%
G_{e}h/(2e^{2})=0.4295-0.2112\log (T$ /$\Delta )+0.0050[\log (T$ /$\Delta
)]^{2}$, with a mean square deviation $6\times 10^{-5}$.

\section{ Summary and discussion}

We have calculated the conductance for a quantum dot embedded in a quantum
wire or side coupled to it as a function of gate voltage and temperature..
We have considered moderate values of $U$ which have not been studied
before. For these values of $U,$ the perturbative method used is quite
accurate. For small coupling of the dot to the wire, the conductance is
determined by an integral of the density of states at the dot times the
derivative of the Fermi function. As the temperature is increased, for $%
U\geq 4\Delta $ we find that the conductance as a function of the gate
voltage changes from a typical shape with one extremum dominated by the
Kondo peak in the spectral density, to another with two extrema,
corresponding to the Coulomb blockade regime. To our knowledge, this
crossover has not been reported previously.

For $U=3\Delta ,$ we find that the conductance can be described as linear in 
$\log T$ as a very good approximation over more than an order of magnitude
in $T.$ This behavior is rather unexpected and difficult to explain in
simple terms, since different energy scales in the problem are of the same
magnitude.

\section{Acknowledgments}

One of us (A.A.A.) wants to thank V. Zlati\'{c} and E. Anda for useful
discussions. The authors are partially supported by CONICET. This work was
sponsored by PICT 03-03833 from ANPCyT and PIP 4952/96 from CONICET.


\begin{references}
\bibitem{note}  The expression $T_{K}=\sqrt{2U\Delta }\exp ^{\pi \varepsilon
_{d}(\varepsilon _{d}+U)/2U\Delta }$ is valid only for very small $\Delta $.
\end{references}
\end{document}